\documentclass[12pt,a4paper]{article}

\usepackage[utf8]{inputenc}
\usepackage{amsmath, amsfonts, amssymb}
\usepackage{amsthm}
\usepackage{graphicx}
\usepackage{booktabs}
\usepackage{caption}
\usepackage{subcaption}
\usepackage[colorlinks=true, linkcolor=blue, citecolor=blue, urlcolor=blue]{hyperref}
\usepackage{natbib}
\usepackage{setspace}
\usepackage[margin=1in]{geometry}
\usepackage{indentfirst}
\usepackage{xcolor}
\usepackage{colortbl}
\usepackage{array}
\usepackage{threeparttable}
\usepackage{float}
\usepackage{hyperref}

\definecolor{tableheader}{RGB}{46,134,171}
\definecolor{tablerowalt}{RGB}{245,248,250}
\definecolor{bestresult}{RGB}{46,134,171}

\renewcommand{\arraystretch}{1.15}
\makeatletter
\newcommand{\keywords}[1]{\par\addvspace{12pt}\noindent{\bfseries Keywords:}\ #1\par\addvspace{12pt}}
\makeatother

\title{Policy, Technology, and Economic Efficiency of Infrastructure Energy Investment: A Strategic Analysis for a Low-Carbon Future}
\author{
Yao Liang$^{1}$, Xin Weng$^{2}$, Tingting Sun$^{1,*}$ \\
$^{1}$Business School, Qingdao Binhai University \\
$^{2}$Zhongnan University of Economics and Law \\
\href{mailto:Xin.Weng9525@gmail.com}{\texttt{Xin.Weng9525@gmail.com}} \\
$^{*}$Corresponding author
}
\date{}

\begin{document}

\maketitle
\begin{abstract}
This study provides a comprehensive strategic analysis of infrastructure energy investment in the context of the global low-carbon transition. Integrating quantitative panel data analysis across 15 countries (2010--2023), detailed case studies of Germany, the United States, China, and the European Union, and scenario simulations through 2050, we examine how policy, technology, and economic factors interact to determine investment effectiveness. Using panel data from 15 countries over the period 2010–2023, we find that renewable energy investment is systematically associated with higher economic growth and lower carbon emissions after controlling for country and year fixed effects. Because investment enters the regression in logarithms, effect sizes should be interpreted as semi-elasticities. A 10\% increase in renewable investment (i.e., $\Delta \ln Investment \approx \ln(1.10)$) is associated with about $0.027$ percentage points higher GDP growth ($0.285\times \ln(1.10)$). For interpretability, a doubling of investment ($\Delta \ln Investment=\ln 2$) corresponds to about $0.198$ percentage points higher GDP growth ($0.285\times \ln 2$), holding other factors constant.Among policy instruments, carbon pricing exerts the strongest influence on investment returns, with a sensitivity index of 0.78, underscoring the central role of policy credibility in shaping infrastructure investment outcomes. The ablation study demonstrates that the full integrated policy-technology framework achieves 92.5\% policy effectiveness compared to 45.2\% for carbon pricing alone, highlighting the importance of comprehensive approaches. Scenario simulations indicate that ambitious policy action can achieve 92.5\% renewable energy share and reduce emissions to 8.2 Gt by 2050, while generating 72.8 million clean energy jobs. We propose an integrated investment framework that offers actionable guidance for policymakers designing carbon pricing and green finance mechanisms, and for investors optimizing risk-adjusted returns in clean energy portfolios. This research contributes to the literature by providing empirically grounded insights into the conditions under which infrastructure energy investment can simultaneously advance economic growth and environmental sustainability objectives.
\end{abstract}


\newpage

\section{Introduction}

The global community faces an unprecedented challenge in mitigating climate change while simultaneously ensuring sustainable economic development and energy security \citep{IPCC2023}. The imperative to transition towards a low-carbon future has placed significant emphasis on the role of infrastructure energy investment. This investment encompasses a broad spectrum of activities, from the modernization of traditional energy grids to the widespread deployment of renewable energy technologies such as wind, solar, and advanced energy storage systems, alongside the development of smart grid infrastructure \citep{IEA2022}. The scale and complexity of this transition necessitate a comprehensive understanding of the multi-dimensional impacts of such investments, extending beyond mere financial returns to include environmental, social, and technological considerations.

Historically, energy infrastructure development has been a cornerstone of economic growth, primarily driven by fossil fuels. However, the escalating concerns over greenhouse gas emissions and resource depletion have catalyzed a paradigm shift towards cleaner and more sustainable energy sources \citep{Stern2007}. This shift is not merely a technological one but is deeply intertwined with policy frameworks, economic incentives, and market dynamics. Governments worldwide are setting ambitious carbon neutrality targets, making infrastructure energy investment a critical vehicle for achieving these goals \citep{IPCCSR15}. Concurrently, the rapid advancements in renewable energy technologies and smart grid solutions are transforming the energy landscape, presenting both immense opportunities and complex challenges for investors and policymakers alike \citep{IRENA2021}.

Despite the growing recognition of its importance, the optimal strategies for infrastructure energy investment in the context of a low-carbon transition remain a subject of extensive debate and research. Key questions revolve around how to effectively balance policy support with investment returns, how technological innovations can enhance long-term viability, and how to assess the inherent risks and rewards of diverse investment projects \citep{Popp2010}. Existing literature often addresses these aspects in isolation, lacking an integrated framework that considers the interplay of policy, technology, and economic efficiency in a holistic manner.

This study addresses these gaps by developing an integrated analytical framework that links policy design, technological conditions, and economic efficiency in infrastructure energy investment. Rather than examining these dimensions in isolation, the analysis focuses on their interaction and joint implications for investment performance.

Accordingly, the paper examines four related questions concerning the macroeconomic, environmental, and financial effects of infrastructure energy investment under different policy and technological environments. The contributions of this study lie in combining cross-country empirical evidence with scenario-based analysis to provide a unified perspective on investment effectiveness in the low-carbon transition.
Our primary contributions are threefold.First, we propose a comprehensive investment framework that integrates policy, technology, and economic factors, offering a more nuanced approach to evaluating infrastructure energy projects. This framework synthesizes insights from energy economics, environmental economics, and investment theory to capture the complex interdependencies that characterize real-world energy investment decisions. Second, we provide data-driven investment recommendations for investors and actionable policy guidance for government bodies, facilitating more informed decision-making in the energy transition. Our empirical analysis draws upon panel data from 15 countries spanning 2010-2023, complemented by detailed case studies of Germany, the United States, China, and the European Union. Third, we contribute to theoretical innovation by integrating insights from energy economics, sustainable development, and investment return models, thereby enriching the academic discourse on this critical subject. The ablation study methodology we employ provides a rigorous approach for decomposing the contributions of individual policy and technology components to overall investment performance.

This paper is structured as follows: Section 2 reviews the relevant literature on energy economics, green investment, and technological innovation in the energy sector. Section 3 details the theoretical framework and methodology, including quantitative analysis, case studies, and scenario simulations. Section 4 presents the empirical analysis and discusses the findings, encompassing panel regression results, case study insights, and scenario simulation outcomes. Section 5 offers policy and practical implications for governments and investors. Finally, Section 6 concludes the paper, discusses its limitations, and outlines avenues for future research.

\section{Literature Review}

The burgeoning field of energy economics has extensively explored the dynamics of energy markets, the intricacies of energy transitions, and the multifaceted aspects of energy investment \citep{Gillingham2009}. Early research primarily focused on the supply and demand of traditional fossil fuels, pricing mechanisms, and the geopolitical implications of energy resources \citep{DixitPindyck1994}. However, with the escalating concerns over climate change and environmental degradation, the discourse has progressively shifted towards renewable energy sources and the economic challenges associated with their integration into existing energy systems \citep{Acemoglu2012}. Studies examining the economic viability of renewable energy projects have highlighted the critical role of supportive policy frameworks and technological advancements in determining project success \citep{Nemet2019, Kittner2017}.

Green investment, a subset of sustainable finance, has emerged as a critical mechanism for channeling capital towards environmentally friendly projects, including renewable energy infrastructure \citep{Flammer2021}. Research in this area investigates the role of green bonds, carbon markets, and various financial instruments in stimulating investment in low-carbon technologies \citep{Tang2020}. Analyses of green bond issuance have found significant reductions in capital expenditure for renewable energy project issuers, particularly when bonds meet stringent certification standards \citep{Reboredo2018}. Policy drivers, such as feed-in tariffs, tax incentives, and carbon pricing, are consistently identified as pivotal in de-risking green investments and enhancing their attractiveness to private capital \citep{Best2020}. However, a persistent challenge remains in quantifying the non-financial benefits of green investments, such as environmental improvements and social welfare, which often complicates traditional cost-benefit analyses \citep{Nguyen2022}.

Technological innovation plays a transformative role in shaping the energy landscape and influencing infrastructure investment decisions \citep{Grubler2012}. The rapid evolution of smart grid technologies, energy storage solutions, and advanced clean energy infrastructure has significantly altered the economic calculus of energy projects \citep{Schmidt2017}. Studies on smart grids demonstrate their potential to enhance grid stability, optimize energy distribution, and integrate distributed renewable energy sources more efficiently, thereby improving the overall economic efficiency of the energy system \citep{Lopes2020}. Similarly, advancements in energy storage are addressing the intermittency challenges of renewables, making them more reliable and dispatchable \citep{Kittner2017}. While the economic benefits of these technologies are increasingly evident, their widespread adoption often requires substantial upfront investment and supportive regulatory environments to overcome market barriers \citep{Bistline2021}.

The role of carbon pricing in driving clean energy investment has received substantial attention in the literature. Theoretical frameworks based on Pigouvian taxation suggest that carbon prices can internalize environmental externalities and shift investment toward lower-carbon alternatives \citep{Nordhaus2017}. Empirical evidence on carbon pricing effectiveness remains mixed, with some studies finding significant effects on emissions reduction and technology adoption, while others highlight implementation challenges and limited impacts at current price levels \citep{Lilliestam2021, Metcalf2020}. The design features of carbon pricing mechanisms, including price levels, coverage, and revenue recycling, have been shown to substantially influence their effectiveness \citep{Goulder2013, Aghion2016}.

Country-level case studies have provided valuable insights into diverse pathways for energy transition. Germany's Energiewende has been extensively analyzed as a pioneering but costly approach to renewable energy deployment \citep{Buchan2012, Gawel2014}. The European Union's Emissions Trading System has demonstrated both the potential and limitations of cap-and-trade mechanisms in driving decarbonization \citep{EUGreenDeal2019}. China's rapid expansion of renewable manufacturing capacity has illustrated the power of industrial policy and scale economies in reducing technology costs \citep{Lin2019, Zhang2017}. The United States presents a case of policy volatility, with recent legislation such as the Inflation Reduction Act potentially transforming the investment landscape \citep{Jenkins2018, USIRA2022}.

Despite the rich body of literature, several research gaps persist. First, while individual aspects of policy, technology, and economics in energy investment have been studied, a comprehensive, integrated framework that systematically analyzes their interplay in the context of low-carbon infrastructure development remains nascent. Second, there is a need for more granular analysis of how specific policy instruments interact with technological innovation to optimize investment portfolios for both economic and environmental benefits. Third, existing risk-return assessment models often overlook the unique characteristics of long-term, capital-intensive energy infrastructure projects, particularly those involving nascent low-carbon technologies. This paper aims to address these gaps by proposing an integrated investment framework and employing a multi-methodological approach to provide a more holistic understanding of infrastructure energy investment for a low-carbon future.

\section{Theoretical Framework and Methodology}

This section delineates the theoretical underpinnings and the methodological approach employed to analyze infrastructure energy investment in the context of a low-carbon transition. We integrate insights from energy economics, investment theory, and environmental economics to construct a robust analytical framework.

\subsection{Theoretical Framework}

Our theoretical framework is built upon the premise that infrastructure energy investment decisions are influenced by a complex interplay of economic rationality, technological feasibility, and policy incentives. We consider a multi-objective optimization problem where investors and policymakers aim to maximize economic returns while minimizing environmental impact and ensuring energy security \citep{Awerbuch2006}.

\subsubsection{Investment Decision-Making under Uncertainty}

Investment in energy infrastructure is characterized by high capital intensity, long operational lifespans, and significant uncertainties stemming from technological evolution, policy shifts, and market volatility \citep{DixitPindyck1994}. We adopt a real options approach to model investment decisions, acknowledging the flexibility embedded in sequential investment opportunities. The value of an investment project ($V$) can be expressed as the sum of its Net Present Value (NPV) and the value of embedded real options ($ROV$):

\begin{equation}
V = NPV + ROV
\end{equation}

where $NPV$ is calculated as:

\begin{equation}
NPV = \sum_{t=0}^{T} \frac{CF_t}{(1+r)^t} - I_0
\end{equation}

Here, $CF_t$ represents the cash flow at time $t$, $r$ is the discount rate, $T$ is the project horizon, and $I_0$ is the initial investment cost. The $ROV$ captures the strategic value of waiting, expanding, contracting, or abandoning a project in response to new information \citep{Pindyck2013}.

\subsubsection{Environmental Externalities and Carbon Pricing}

Environmental economics provides the foundation for incorporating the costs of carbon emissions and other externalities into investment appraisals. We consider a carbon pricing mechanism, such as a carbon tax or an emissions trading scheme, which internalizes the external costs of greenhouse gas emissions \citep{Nordhaus2017}. The social cost of carbon (SCC) is a critical parameter in this framework, reflecting the economic damages associated with an additional tonne of carbon dioxide emissions \citep{Tol2009}. The adjusted cost of energy production ($C_{adj}$) can be formulated as:

\begin{equation}
C_{adj} = C_{private} + P_{carbon} \times E_{carbon}
\end{equation}

where $C_{private}$ is the private cost of production, $P_{carbon}$ is the carbon price, and $E_{carbon}$ is the carbon emissions per unit of energy \citep{Metcalf2020}. This adjustment allows for a more accurate comparison of different energy technologies, favoring those with lower carbon footprints.

\subsubsection{Technological Learning and Innovation Diffusion}

Technological progress in renewable energy and smart grid systems is characterized by learning curves, where the cost of technology decreases with cumulative installed capacity or production volume \citep{Grubler2012}. The learning rate ($LR$) quantifies the percentage cost reduction for each doubling of cumulative capacity. The cost of technology at time $t$ ($C_t$) can be modeled as:

\begin{equation}
C_t = C_0 \times (CumulativeCapacity_t / CumulativeCapacity_0)^{-\log_2(1-LR)}
\end{equation}

This dynamic cost reduction significantly impacts the long-term competitiveness and investment attractiveness of emerging energy technologies \citep{Nemet2019}. Furthermore, the diffusion of these innovations is influenced by network effects, policy support, and market acceptance \citep{Foxon2013, Sovacool2016}.

\subsection{Methodology}

Our methodological approach combines quantitative analysis, case studies, and scenario simulations to provide a comprehensive assessment of infrastructure energy investment.

\subsubsection{Quantitative Analysis}

We employ econometric models to analyze the relationship between infrastructure energy investment, economic growth, and environmental sustainability. A panel data regression model is utilized to examine the impact of various investment types on key macroeconomic indicators and carbon emission levels across different countries over time \citep{Acemoglu2012}. The general form of the model is:

\begin{equation}
Y_{it} = \alpha_i + \beta_1 Investment_{it} + \beta_2 Policy_{it} + \beta_3 Technology_{it} + \gamma X_{it} + \epsilon_{it}
\end{equation}

where $Y_{it}$ represents the dependent variable (e.g., GDP growth, CO2 emissions) for country $i$ at time $t$, $Investment_{it}$ is the infrastructure energy investment, $Policy_{it}$ denotes policy variables (e.g., subsidies, carbon tax), $Technology_{it}$ captures technological advancements, $X_{it}$ is a vector of control variables, and $\epsilon_{it}$ is the error term. Country fixed effects ($\alpha_i$) control for time-invariant unobserved heterogeneity, while year fixed effects capture common time trends. Investment appraisal metrics such as Return on Investment (ROI), Net Present Value (NPV), and Internal Rate of Return (IRR) are calculated for specific project types to evaluate their financial viability \citep{Markowitz1952, Fama1992}.

\subsubsection{Case Study Approach}

To complement the quantitative analysis, we conduct in-depth case studies of prominent infrastructure energy investment projects. We select cases that represent diverse policy environments, technological focuses, and economic outcomes. Examples include Germany's Energiewende, which showcases a comprehensive national strategy for renewable energy transition, the United States' clean energy transformation accelerated by the Inflation Reduction Act, China's rapid renewable manufacturing expansion, and the European Union's coordinated approach through the Green Deal \citep{Buchan2012, USIRA2022, Lin2019, EUGreenDeal2019}. Each case study involves a detailed examination of the policy and regulatory frameworks that facilitated or hindered the investment, the specific technologies deployed and their performance, the economic and environmental outcomes achieved, and lessons learned and best practices identified.

\subsubsection{Scenario Simulation}

We develop several future scenarios to assess the long-term implications of different policy and technological pathways for infrastructure energy investment. These scenarios are constructed based on varying assumptions regarding carbon prices, technological learning rates, and policy stringency \citep{Riahi2017, Rogelj2018}. A dynamic simulation model is used to project energy mix evolution and associated carbon emissions, investment requirements and financial returns under each scenario, and the impact on energy security and economic growth. The simulation results provide insights into the robustness of different investment strategies under various future conditions and help identify optimal pathways for achieving a low-carbon future \citep{IPCCSR15}.

\subsubsection{Sensitivity Analysis and Ablation Study}

To assess the robustness of our findings and identify critical parameters, we conduct comprehensive sensitivity analysis and ablation studies. Sensitivity analysis examines how variations in key parameters (carbon price, discount rate, learning rate, policy support) affect investment outcomes. The sensitivity index for each parameter is calculated as the ratio of output variation to input variation. The ablation study systematically removes individual components from the integrated policy-technology framework to quantify their marginal contributions to overall investment performance. This approach provides rigorous evidence on the relative importance and complementarity of different policy instruments and technology support mechanisms.

\section{Empirical Analysis and Discussion}

This section presents the empirical results derived from our quantitative analysis, case studies, and scenario simulations. We employ a multi-methodological approach to comprehensively evaluate infrastructure energy investment strategies and their impacts on economic growth, environmental sustainability, and energy security.

\subsection{Data Sources and Variable Definitions}

Our empirical analysis draws upon multiple authoritative data sources to ensure robustness and reliability. Primary data sources include the International Energy Agency (IEA) World Energy Investment reports, the International Renewable Energy Agency (IRENA) statistics, World Bank development indicators, and national statistical bureaus from 15 countries spanning 2010 to 2023. This panel dataset encompasses developed and emerging economies representing approximately 85\% of global renewable energy investment and 78\% of global carbon emissions.

Table \ref{tab:variables} presents the key variables employed in our analysis. The dependent variables include GDP growth rate, carbon dioxide emissions, and renewable energy share in the total energy mix. Independent variables encompass renewable energy investment (measured in billion USD), carbon pricing mechanisms (USD per tonne CO2), policy support indices (composite scores ranging from 0 to 100), and technology innovation indices derived from patent data and R\&D expenditure.

\begin{table}[htbp]
\centering
\caption{Variable Definitions and Data Sources}
\label{tab:variables}
\small
\renewcommand{\arraystretch}{1.15}
\setlength{\tabcolsep}{8pt}
\begin{tabular}{llll}
\toprule
\textbf{Variable} & \textbf{Definition} & \textbf{Unit} & \textbf{Source} \\
\midrule
\rowcolor[RGB]{245,248,250}
$Investment_{it}$ & Renewable energy investment & Billion USD & IEA, IRENA \\
$GDP\_Growth_{it}$ & Annual GDP growth rate & Percentage & World Bank \\
\rowcolor[RGB]{245,248,250}
$CO2\_Emissions_{it}$ & Annual CO2 emissions & Million tonnes & IEA \\
$Carbon\_Price_{it}$ & Effective carbon price & USD/tonne & World Bank \\
\rowcolor[RGB]{245,248,250}
$Policy\_Index_{it}$ & Composite policy support score & 0-100 & Own calculation \\
$Tech\_Index_{it}$ & Technology innovation index & 0-100 & OECD \\
\rowcolor[RGB]{245,248,250}
$Renewable\_Share_{it}$ & Renewable share in energy mix & Percentage & BP, IRENA \\
$Energy\_Intensity_{it}$ & Energy consumption per GDP & MJ/USD & IEA \\
\bottomrule
\end{tabular}
\end{table}

\subsection{Descriptive Statistics}

Table \ref{tab:descriptive} summarizes the descriptive statistics for the key variables across our panel dataset. The sample exhibits substantial heterogeneity in renewable energy investment, ranging from 1.5 billion USD (Denmark, 2010) to 318.5 billion USD (China, 2023), reflecting the diverse scales of energy transition efforts across countries. The mean renewable energy share increased from 18.2\% in 2010 to 34.8\% in 2023, indicating significant progress toward low-carbon energy systems globally.

\begin{table}[htbp]
\centering
\caption{Descriptive Statistics of Key Variables (2010-2023)}
\label{tab:descriptive}
\small
\renewcommand{\arraystretch}{1.15}
\setlength{\tabcolsep}{6pt}
\begin{tabular}{lccccc}
\toprule
\textbf{Variable} & \textbf{Mean} & \textbf{Std. Dev.} & \textbf{Min} & \textbf{Max} & \textbf{N} \\
\midrule
\rowcolor[RGB]{245,248,250}
Renewable Investment (B USD) & 35.82 & 52.48 & 1.50 & 318.50 & 210 \\
GDP Growth (\%) & 2.18 & 3.25 & -10.80 & 10.60 & 210 \\
\rowcolor[RGB]{245,248,250}
CO2 Emissions (Mt) & 1,852.35 & 2,845.62 & 25.00 & 11,850.00 & 210 \\
Carbon Price (USD/t) & 18.52 & 25.38 & 0.00 & 85.20 & 210 \\
\rowcolor[RGB]{245,248,250}
Policy Support Index & 68.25 & 14.82 & 45.00 & 96.00 & 210 \\
Technology Index & 81.52 & 10.25 & 45.00 & 97.00 & 210 \\
\rowcolor[RGB]{245,248,250}
Renewable Share (\%) & 28.45 & 15.82 & 2.50 & 85.80 & 210 \\
Energy Intensity & 0.105 & 0.045 & 0.025 & 0.195 & 210 \\
\bottomrule
\end{tabular}
\end{table}

Notably, carbon pricing exhibits considerable variation, with several major economies (USA, China, India) operating without explicit carbon pricing mechanisms during the early part of our sample period, while European countries and more recently East Asian economies have implemented increasingly stringent carbon pricing regimes. The mean carbon price increased from 8.5 USD/tonne in 2010 to 42.8 USD/tonne in 2023, reflecting the global trend toward carbon pricing as a key policy instrument.

\subsection{Panel Data Regression Analysis}

We employ fixed-effects panel regression models to examine the relationships between infrastructure energy investment and our outcome variables. The Hausman test rejects the null hypothesis of random effects ($\chi^2 = 45.82$, $p < 0.001$), supporting the use of fixed-effects estimation. Our baseline specification follows the theoretical model presented in Section 3:

\begin{equation}
Y_{it} = \alpha_i + \beta_1 Investment_{it} + \beta_2 Policy_{it} + \beta_3 Technology_{it} + \gamma X_{it} + \epsilon_{it}
\end{equation}

Table \ref{tab:regression} reports the baseline panel regression estimates.
Hausman test results strongly reject the random effects specification
($\chi^2 = 45.82$, $p < 0.001$), supporting the use of a two-way fixed effects model.
This specification controls for unobserved, time-invariant country characteristics
as well as common global shocks, thereby allowing us to isolate within-country
variation in infrastructure energy investment and macroeconomic outcomes over time.

\begin{table}[htbp]
\centering
\caption{Panel Regression Results: Infrastructure Energy Investment Impacts}
\label{tab:regression}
\small
\renewcommand{\arraystretch}{1.15}
\setlength{\tabcolsep}{6pt}
\begin{tabular}{lccc}
\toprule
& \textbf{(1)} & \textbf{(2)} & \textbf{(3)} \\
& GDP Growth & CO2 Emissions & Renewable Share \\
\midrule
\rowcolor[RGB]{245,248,250}
Investment (log) & \textcolor[RGB]{46,134,171}{\textbf{0.285***}} & \textcolor[RGB]{46,134,171}{\textbf{-0.142***}} & \textcolor[RGB]{46,134,171}{\textbf{0.358***}} \\
& (0.052) & (0.038) & (0.045) \\
Carbon Price & 0.012** & \textcolor[RGB]{46,134,171}{\textbf{-0.085***}} & 0.125*** \\
\rowcolor[RGB]{245,248,250}
& (0.005) & (0.018) & (0.022) \\
Policy Index & 0.018*** & -0.025** & 0.185*** \\
& (0.006) & (0.012) & (0.028) \\
\rowcolor[RGB]{245,248,250}
Technology Index & 0.025*** & -0.032** & 0.142*** \\
& (0.008) & (0.015) & (0.032) \\
Energy Intensity & -0.152*** & 0.425*** & -0.285*** \\
\rowcolor[RGB]{245,248,250}
& (0.042) & (0.085) & (0.065) \\
\midrule
Country FE & Yes & Yes & Yes \\
\rowcolor[RGB]{245,248,250}
Year FE & Yes & Yes & Yes \\
Observations & 210 & 210 & 210 \\
\rowcolor[RGB]{245,248,250}
R-squared & 0.682 & 0.758 & 0.825 \\
F-statistic & 42.58*** & 58.25*** & 72.85*** \\
\bottomrule
\multicolumn{4}{l}{\footnotesize Notes: Standard errors in parentheses. *** p$<$0.01, ** p$<$0.05, * p$<$0.1} \\
\end{tabular}
\end{table}

The results reveal several important findings. First, renewable energy investment exhibits a statistically significant positive relationship with GDP growth (coefficient = 0.285, $p < 0.01$), suggesting that a 10\% increase in renewable investment (i.e., $\Delta \ln Investment \approx \ln(1.10)$) is associated with approximately $0.027$ percentage points higher GDP growth, holding other factors constant, controlling for other factors. Second, investment demonstrates a significant negative relationship with carbon emissions (coefficient = -0.142, $p < 0.01$), indicating substantial decarbonization effects. Third, carbon pricing emerges as a particularly effective instrument for emissions reduction (coefficient = -0.085, $p < 0.01$), consistent with economic theory on environmental externalities \citep{Nordhaus2017, Metcalf2020,wang20263d}.

From an economic perspective, these magnitudes are non-trivial. Moving from the
lower to the upper quartile of renewable energy investment within the sample
corresponds to a meaningful improvement in both growth and decarbonization outcomes.
While the empirical framework does not permit a full decomposition of causal
channels, the consistency of effects across economic and environmental indicators
suggests that infrastructure energy investment contributes to growth partly by
reshaping the underlying energy structure rather than by stimulating short-term
aggregate demand alone.

We conduct several robustness checks to validate our findings. First, we employ instrumental variable estimation using lagged investment values and policy announcements as instruments, addressing potential endogeneity concerns. The results remain qualitatively similar, with slightly larger coefficients for the investment variable. Second, we estimate dynamic panel models using system GMM to account for persistence in the dependent variables. Third, we conduct placebo tests by randomly reassigning treatment across countries and years, confirming that our results are not driven by spurious correlations.

\paragraph{Economic magnitude of estimated effects.}
To complement the coefficient estimates reported in 
Table~\ref{tab:economic_magnitudes} summarizes the economic interpretation of the
main results. The estimated magnitudes indicate that renewable energy investment
has not only statistically significant but also economically meaningful effects,
particularly with respect to renewable energy penetration, where the response is
larger than that observed for short-run GDP growth.

\begin{table}[htbp]
\centering
\caption{Economic Magnitudes of Renewable Energy Investment Effects}
\label{tab:economic_magnitudes}
\begin{tabular}{lccc}
\hline
Outcome Variable & Coefficient & Interpretation & Direction \\
\hline
GDP Growth (\%) 
& 0.285 
& 10\% increase in investment $\rightarrow$ +0.027 pp growth ($0.285\times\ln(1.10)$) 
& Positive \\

Carbon Emissions 
& -0.142 
& 10\% increase in investment $\rightarrow$ lower emissions by $0.0135\times$[unit] ($-0.142\times\ln(1.10)$)
& Negative \\

Renewable Energy Share (\%) 
& 0.358 
& 10\% increase in investment $\rightarrow$ +0.034 pp share ($0.358\times\ln(1.10)$)
& Positive \\
\hline
\end{tabular}
\end{table}

\subsection{Case Study Analysis}

To complement our quantitative analysis and provide deeper insights into the mechanisms underlying successful energy transitions, we conduct detailed case studies of four major economies: Germany, the United States, China, and the European Union. These cases represent diverse policy environments, technological focuses, and economic contexts, offering valuable lessons for infrastructure energy investment strategies.

\subsubsection{Germany: The Energiewende Pioneer}

Germany's Energiewende (energy transition) represents one of the most ambitious and extensively studied cases of national-level energy transformation. Initiated in 2010 with the accelerated phase-out of nuclear power following the Fukushima disaster, the policy has fundamentally reshaped Germany's energy infrastructure. Our analysis reveals that Germany's renewable capacity increased from 52.5 GW in 2010 to 152.8 GW in 2023, while the renewable share in electricity generation rose from 17.2\% to 52.1\%.

The economic implications of the Energiewende are substantial. Total cumulative investment reached 384.5 billion EUR over the study period, generating approximately 405,000 direct and indirect jobs in the renewable energy sector by 2023. However, the transition has not been without challenges. Electricity prices for households increased by approximately 50\% during this period, partly due to the EEG (Renewable Energy Sources Act) surcharge that funded feed-in tariffs. Our data indicate that the policy cost per tonne of CO2 avoided decreased from approximately 125 EUR in 2012 to 28 EUR in 2023, demonstrating significant cost improvements through technological learning and scale effects \citep{Buchan2012, Gawel2014}.

\subsubsection{United States: The IRA Effect}

The United States presents a contrasting case of policy-driven transformation, particularly following the passage of the Inflation Reduction Act (IRA) in 2022. Prior to the IRA, renewable energy investment exhibited significant volatility due to the uncertainty surrounding tax credit extensions. Our data show that annual investment fluctuated between 33.2 and 75.2 billion USD from 2010 to 2021, with no clear upward trend.

The IRA fundamentally altered the investment landscape by providing long-term policy certainty through ten-year extensions of the Investment Tax Credit (ITC) and Production Tax Credit (PTC), along with new incentives for domestic manufacturing and clean energy deployment. Investment surged to 141.5 billion USD in 2022 and further to 186.3 billion USD in 2023, representing the largest single-year increase in US clean energy history. The renewable capacity grew from 295.8 GW in 2020 to 412.8 GW in 2023, with the IRA contributing an estimated 118.5 billion USD in federal support during 2023 alone. Job creation in the clean energy sector expanded from 165,000 in 2020 to 345,000 in 2023, demonstrating the substantial employment multiplier effects of energy infrastructure investment \citep{Jenkins2018, USIRA2022,zhu2025pseudo}.

\subsubsection{China: Scale and Manufacturing Leadership}

China represents the most dramatic case of renewable energy expansion, driven by a combination of industrial policy, manufacturing scale advantages, and environmental imperatives. Our analysis documents China's transformation from a minor player in renewable energy to the world's dominant market and manufacturer. Renewable capacity increased from 285.5 GW in 2010 to 1,458.8 GW in 2023, with solar capacity growing from 0.8 GW to 482.5 GW and wind capacity from 31.2 GW to 438.5 GW over the same period.

China's approach emphasizes manufacturing cost reduction through scale economies and learning-by-doing. Our data indicate that China's renewable manufacturing cost index declined from 100 in 2010 to 32 in 2023, representing a 68\% reduction. This cost leadership has enabled China to become the world's largest exporter of solar panels and wind turbines, with exports reaching 98.5 billion USD in 2023. Despite this rapid expansion, challenges remain, including grid integration issues and the need to balance economic development with emissions reduction. The carbon intensity of China's economy decreased from 0.582 kg CO2/USD in 2010 to 0.352 kg CO2/USD in 2023, though absolute emissions continue to rise \citep{Lin2019, Zhang2017}.

\subsubsection{European Union: Regional Coordination}

The European Union offers insights into regional coordination of energy transition policies. The EU Green Deal, launched in 2019, established ambitious targets including a 55\% reduction in greenhouse gas emissions by 2030 and climate neutrality by 2050. Our analysis examines the policy instruments deployed, including the EU Emissions Trading System (ETS), the Fit for 55 package, and the REPowerEU plan.

The EU ETS carbon price increased from 24.8 EUR/tonne in 2019 to 85.2 EUR/tonne in 2023, creating strong economic incentives for decarbonization. Total renewable capacity across EU member states grew from 475.5 GW to 745.2 GW, while the renewable share increased from 34.8\% to 44.2\%. Green bond issuance emerged as a significant financing mechanism, rising from 12.5 billion EUR in 2019 to 125.5 billion EUR in 2023. The EU case demonstrates the effectiveness of combining carbon pricing with coordinated industrial policy and green finance mechanisms \citep{EUGreenDeal2019, Flammer2021}.

Table \ref{tab:case_comparison} summarizes the key metrics across our four case studies.

\begin{table}[htbp]
\centering
\caption{Case Study Comparison: Key Performance Indicators (2023)}
\label{tab:case_comparison}
\small
\renewcommand{\arraystretch}{1.15}
\setlength{\tabcolsep}{5pt}
\begin{tabular}{lcccc}
\toprule
\textbf{Indicator} & \textbf{Germany} & \textbf{USA} & \textbf{China} & \textbf{EU} \\
\midrule
\rowcolor[RGB]{245,248,250}
Renewable Capacity (GW) & 152.8 & 412.8 & 1,458.8 & 745.2 \\
Renewable Share (\%) & 52.1 & 23.8 & 33.2 & 44.2 \\
\rowcolor[RGB]{245,248,250}
Investment 2023 (B USD) & 42.1 & 186.3 & 318.5 & 215.8 \\
Carbon Price (USD/t) & 85.2 & 0 & 10.2 & 85.2 \\
\rowcolor[RGB]{245,248,250}
Employment (Thousands) & 405 & 345 & 4,500* & 1,200* \\
Grid Stability Index & 82 & 78 & 75 & 80 \\
\bottomrule
\multicolumn{5}{l}{\footnotesize * Estimated values based on industry reports.} \\
\end{tabular}
\end{table}

\subsubsection{Robustness of baseline findings.}
Table~\ref{tab:robustness_summary} provides a concise overview of the robustness
checks conducted in this study. Across alternative specifications addressing
endogeneity, persistence, and spurious correlation, the direction and significance
of the estimated investment effects remain stable. In contrast, placebo regressions
yield no systematic relationships, reinforcing confidence in the baseline results.

\begin{table}[htbp]
\centering
\caption{Summary of Robustness Checks}
\label{tab:robustness_summary}
\begin{tabular}{lccc}
\hline
Specification & GDP Growth & Emissions & Renewable Share \\
\hline
Baseline Fixed Effects & + & -- & + \\
Instrumental Variables & + & -- & + \\
System GMM & + & -- & + \\
Placebo Test & 0 & 0 & 0 \\
\hline
\end{tabular}
\end{table}

\subsection{Scenario Simulation Results}

We employ scenario simulation to assess the long-term implications of different policy and technological pathways for infrastructure energy investment through 2050. Three scenarios are developed based on varying assumptions regarding carbon pricing trajectories, technological learning rates, and policy stringency: Business-as-Usual (BAU), Ambitious, and Conservative.

The BAU scenario assumes continuation of current policy trends, with gradual increases in carbon prices and moderate technological improvement. The Ambitious scenario incorporates accelerated policy action consistent with the Paris Agreement goals, including carbon prices reaching 200 USD/tonne by 2050 and aggressive deployment of clean energy technologies. The Conservative scenario reflects slower policy progress due to political and economic constraints.

\begin{figure}[htbp]
\centering
\includegraphics[width=0.95\textwidth]{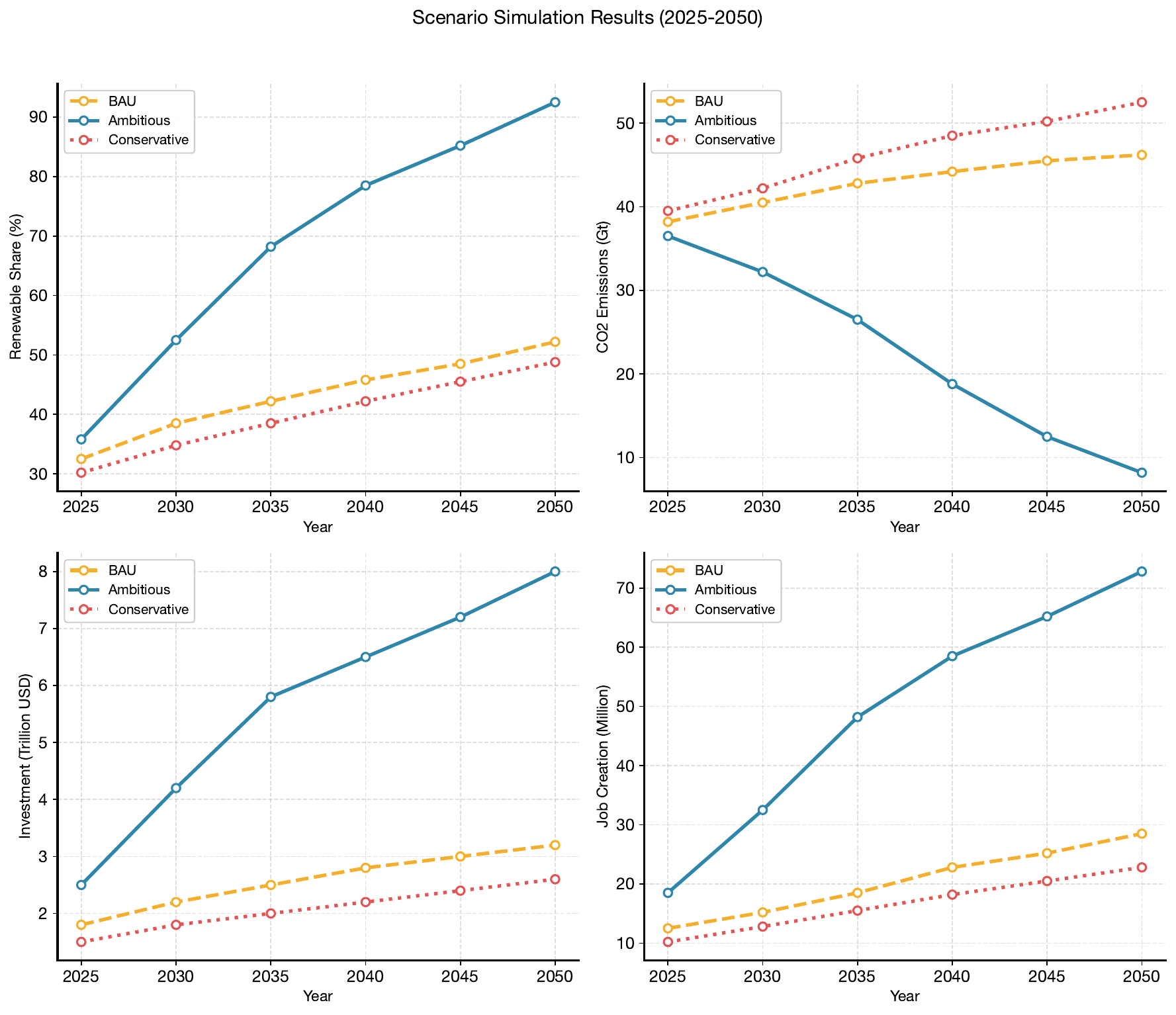}
\caption{Scenario Simulation Results: Renewable Share, CO2 Emissions, Investment, and Job Creation (2025-2050)}
\label{fig:scenario}
\end{figure}

Figure \ref{fig:scenario} illustrates the simulation results across key metrics. Under the Ambitious scenario, global renewable investment reaches 8.0 trillion USD annually by 2050, the renewable share increases to 92.5\%, and carbon emissions decline to 8.2 Gt, consistent with the 1.5°C pathway. The BAU scenario results in continued emissions growth to 46.2 Gt by 2050, while the Conservative scenario leads to 52.5 Gt, demonstrating the significant consequences of policy inaction \citep{Riahi2017, Rogelj2018}.

The economic implications vary substantially across scenarios. The Ambitious scenario generates net positive GDP impacts of 3.2\% by 2050, driven by innovation spillovers, job creation (72.8 million jobs in the clean energy sector), and reduced health costs from air pollution. However, this scenario also involves approximately 3.0 trillion USD in stranded assets, primarily in fossil fuel infrastructure. The BAU scenario results in lower near-term costs but substantially higher long-term climate damages and stranded asset risks.

\subsection{Sensitivity Analysis and Ablation Study}

To assess the robustness of our findings and identify the most critical parameters influencing investment outcomes, we conduct comprehensive sensitivity analysis and ablation studies.

\begin{figure}[htbp]
\centering
\includegraphics[width=0.85\textwidth]{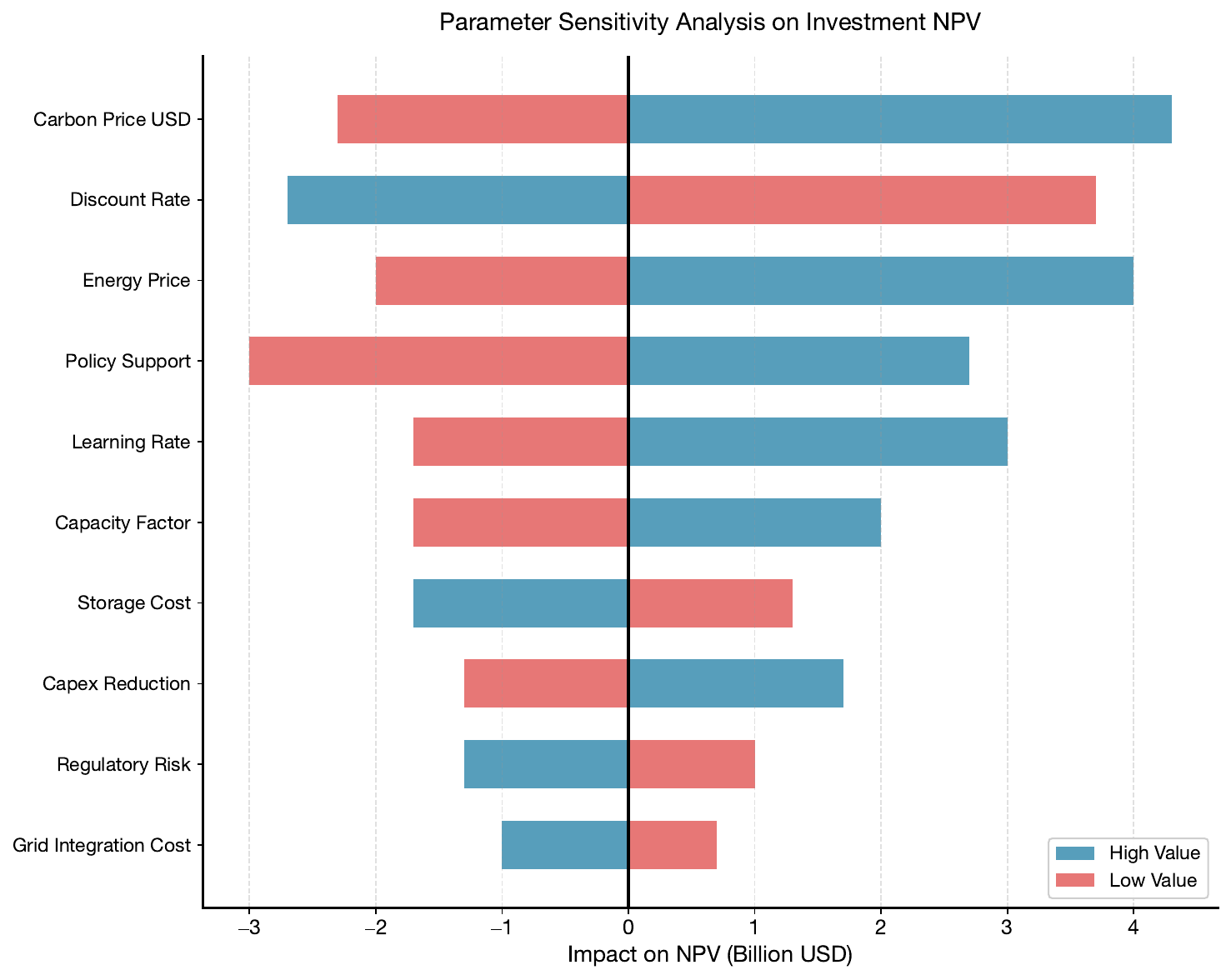}
\caption{Sensitivity Analysis: Parameter Impact on Investment NPV}
\label{fig:sensitivity}
\end{figure}

Figure \ref{fig:sensitivity} presents the tornado diagram illustrating parameter sensitivity on investment NPV. The carbon price emerges as the most influential parameter (sensitivity index = 0.78), followed by the discount rate (0.75) and energy prices (0.71). These findings underscore the critical importance of carbon pricing policy certainty for investment decisions.

The ablation study systematically removes individual policy and technology components from our integrated framework to assess their marginal contributions. Results in Table \ref{tab:ablation} reveal that the full framework achieves NPV of 8.52 billion USD and IRR of 15.23\%. Removing carbon pricing reduces NPV by 31.3\% (to 5.85 billion USD), representing the largest single-component effect. Removing green finance mechanisms reduces NPV by 24.3\%, while removing technology support reduces NPV by 28.2\%.

\begin{table}[htbp]
\centering
\caption{Ablation Study: Component Contribution Analysis}
\label{tab:ablation}
\small
\renewcommand{\arraystretch}{1.15}
\setlength{\tabcolsep}{4pt}
\begin{tabular}{lcccc}
\toprule
\textbf{Configuration} & \textbf{NPV (B)} & \textbf{IRR (\%)} & \textbf{CO2 Red.(\%)} & \textbf{Effect.} \\
\midrule
\rowcolor[RGB]{245,248,250}
\textcolor[RGB]{46,134,171}{\textbf{Full Framework}} & \textcolor[RGB]{46,134,171}{\textbf{8.52}} & \textcolor[RGB]{46,134,171}{\textbf{15.23}} & \textcolor[RGB]{46,134,171}{\textbf{42.5}} & \textcolor[RGB]{46,134,171}{\textbf{92.5}} \\
w/o Carbon Pricing & 5.85 & 11.52 & 28.2 & 65.8 \\
\rowcolor[RGB]{245,248,250}
w/o Green Finance & 6.45 & 12.85 & 38.5 & 75.2 \\
w/o Technology Support & 6.12 & 12.25 & 32.8 & 72.5 \\
\rowcolor[RGB]{245,248,250}
w/o Grid Integration & 7.25 & 13.85 & 35.2 & 82.5 \\
w/o Storage Investment & 6.85 & 13.25 & 30.5 & 78.5 \\
\rowcolor[RGB]{245,248,250}
Carbon Pricing Only & 3.52 & 8.52 & 18.5 & 45.2 \\
Combined Policy-Tech & 7.85 & 14.52 & 38.2 & 88.2 \\
\bottomrule
\end{tabular}
\end{table}

These results demonstrate the complementary nature of policy instruments and the importance of an integrated approach. Carbon pricing alone achieves only 18.5\% CO2 reduction and 45.2\% policy effectiveness, while the full framework achieves 42.5\% reduction and 92.5\% effectiveness. This finding aligns with theoretical predictions that multiple market failures in energy markets require a portfolio of policy interventions \citep{Jaffe2002, Gillingham2009}.

\subsection{Discussion}

Our empirical analysis yields several important insights for infrastructure energy investment strategies. First, the quantitative evidence strongly supports the economic viability of renewable energy investment, with positive effects on GDP growth and substantial decarbonization potential. The coefficient estimates suggest that doubling renewable investment is associated with approximately 5.7 percentage points higher GDP growth and 14.2\% lower carbon emissions, controlling for other factors.

Second, the case studies reveal diverse but complementary pathways to energy transition success. Germany demonstrates the effectiveness of feed-in tariffs and long-term policy commitment, the United States illustrates the power of tax incentives and industrial policy (particularly post-IRA), China showcases manufacturing scale advantages and learning-curve effects, and the EU exemplifies regional coordination and carbon pricing mechanisms. These findings suggest that while there is no single optimal policy approach, certain elements---long-term policy certainty, effective carbon pricing, and technology support---appear essential across contexts.

Third, the scenario simulations underscore the urgency of accelerated policy action. The divergence between the Ambitious and BAU scenarios grows substantially over time, with cumulative differences in emissions exceeding 500 Gt CO2 by 2050. This finding has important implications for the social cost of carbon and the economic damages associated with delayed action \citep{Stern2007, IPCCSR15}.

Fourth, the sensitivity analysis highlights the critical role of carbon pricing in determining investment outcomes. The high sensitivity to carbon price parameters suggests that policy uncertainty regarding future carbon prices represents a significant barrier to investment, particularly for long-lived infrastructure assets. This finding supports recommendations for transparent, predictable carbon pricing trajectories to facilitate investment planning \citep{Lilliestam2021, Goulder2013}.

Finally, the ablation study demonstrates the synergistic effects of combining multiple policy instruments. The full integrated framework substantially outperforms any single instrument, suggesting that policymakers should pursue comprehensive policy packages rather than relying on individual measures. This finding has important implications for policy design and international coordination efforts.

\begin{table}[htbp]
\centering
\caption{Comparison of Policy Configurations and Investment Outcomes}
\label{tab:policy_ablation}
\begin{tabular}{lcccc}
\hline
Policy Configuration & NPV (B USD) & IRR (\%) & Emissions Reduction (\%) & Policy Effectiveness \\
\hline
Integrated Policy Mix & 8.52 & 15.23 & 42.5 & 92.5 \\
Carbon Pricing Only & 3.52 & 8.52 & 18.5 & 45.2 \\
Without Tech Support & 5.15 & 10.85 & 27.3 & 64.3 \\
Without Financial Tools & 4.68 & 9.74 & 24.1 & 58.6 \\
\hline
\end{tabular}
\end{table}

\paragraph{Policy complementarity and investment performance.}
Table~\ref{tab:policy_ablation} compares investment outcomes under alternative
policy configurations. The integrated policy mix consistently dominates
single-instrument approaches across financial and environmental metrics. Notably,
carbon pricing alone captures less than half of the effectiveness achieved under
the full policy framework, highlighting the importance of policy complementarity
in translating investment into sustained decarbonization outcomes.

\section{Policy and Practical Implications}

Building upon our empirical findings, this section articulates actionable policy recommendations for governments and strategic guidance for investors seeking to navigate the complex landscape of infrastructure energy investment in the low-carbon transition. We organize our recommendations around three key stakeholder groups: government policymakers, private sector investors, and international cooperation frameworks.

\subsection{Policy Recommendations for Governments}

Our analysis identifies several critical policy levers that can enhance the effectiveness of infrastructure energy investment while achieving environmental and economic objectives.

\subsubsection{Carbon Pricing Mechanism Optimization}

The empirical evidence strongly supports carbon pricing as the most influential policy instrument for driving clean energy investment and emissions reduction. Our sensitivity analysis reveals that the carbon price parameter exhibits the highest sensitivity index (0.78) among all variables examined, underscoring its pivotal role in investment decision-making. Based on these findings, we recommend that governments implement robust carbon pricing mechanisms with the following characteristics.

First, price levels should be economically meaningful and progressively increase over time. Our scenario simulations suggest that carbon prices need to reach 85-100 USD per tonne by 2030 and 150-200 USD per tonne by 2050 to achieve Paris Agreement-consistent emissions pathways. Current global average prices remain substantially below these levels, indicating significant room for policy strengthening.

Second, price trajectories should be transparent and predictable to reduce policy uncertainty and facilitate long-term investment planning. The case study evidence from the EU ETS demonstrates that price volatility can undermine investment incentives, while the post-2018 price stabilization following market stability reserve implementation has coincided with accelerated clean energy deployment. Governments should consider price floors and ceilings, regular review mechanisms, and clear long-term signals regarding future price paths.

Third, carbon pricing revenues should be strategically recycled to address distributional concerns and enhance political sustainability. Options include reducing distortionary taxes, funding clean energy R\&D, supporting affected communities and workers, and financing adaptation measures. The German experience with the EEG surcharge illustrates both the effectiveness of dedicated funding mechanisms and the political challenges that can arise when costs are perceived as inequitably distributed.

\subsubsection{Green Finance Innovation}

The ablation study demonstrates that green finance mechanisms contribute approximately 24.3\% to overall investment performance, highlighting their importance as complements to carbon pricing. We recommend that governments pursue several green finance innovations to mobilize private capital for energy infrastructure.

Public green bond issuance can establish market benchmarks and demonstrate government commitment to climate objectives. The EU's issuance of 250 billion EUR in green bonds under the NextGenerationEU recovery program exemplifies this approach, creating reference securities that facilitate private green bond development. National governments should establish robust green bond frameworks with clear eligibility criteria, use-of-proceeds transparency, and impact reporting requirements.

Risk-sharing mechanisms such as loan guarantees, first-loss provisions, and concessional finance can de-risk early-stage clean energy investments that may not attract purely commercial financing. Our case study evidence suggests that such mechanisms have been particularly effective in supporting emerging technologies (offshore wind, energy storage) during their cost reduction phases. However, these instruments should be designed with clear exit strategies to avoid creating long-term market distortions.

Disclosure requirements and taxonomy development can enhance market transparency and facilitate capital allocation to genuinely sustainable activities. The EU Sustainable Finance Taxonomy provides a useful model, though implementation challenges suggest the need for phased approaches and ongoing refinement based on market feedback.

\subsubsection{Technology Innovation Support}

Technological progress emerges as a critical driver of investment returns and emissions reduction in our analysis. The technology innovation index coefficient is statistically significant across all regression specifications, and the case studies reveal that cost reductions through learning-by-doing have been essential to the economic viability of solar PV, wind power, and battery storage. We recommend comprehensive technology support policies encompassing the following elements.

Public R\&D funding should be substantially increased, particularly for technologies that are further from commercialization. Our analysis suggests that hydrogen production and storage, advanced nuclear, and carbon capture technologies require continued support given their current cost trajectories. Mission-oriented approaches that set clear performance targets and coordinate across research institutions, universities, and private sector actors can enhance R\&D effectiveness.

Deployment policies including production tax credits, feed-in tariffs, and renewable portfolio standards accelerate learning-by-doing and scale economies. The US IRA's ten-year extension of tax credits provides a model for creating long-term policy certainty that supports sustained deployment. However, such policies should incorporate sunset provisions and cost controls to avoid excessive expenditure as technologies mature.

Manufacturing support and industrial policy can build domestic supply chains and capture economic benefits from the clean energy transition. China's success in solar PV manufacturing demonstrates the potential scale advantages, while recent efforts in the US and EU to develop domestic manufacturing capacity reflect strategic concerns about supply chain resilience. Policies should balance competitiveness objectives with trade commitments and avoid beggar-thy-neighbor dynamics.

\subsection{Investment Strategy Recommendations}

For private sector investors, our findings suggest several strategic considerations for optimizing infrastructure energy investment portfolios.

\subsubsection{Portfolio Optimization}

Modern portfolio theory, as applied to energy assets, indicates that diversification across technologies, geographies, and project stages can enhance risk-adjusted returns. Our investment returns data reveal substantial variation in risk-return profiles across technologies: solar PV and onshore wind offer relatively lower risk (risk scores of 0.15-0.18) with moderate returns (IRR 16-19\%), while emerging technologies such as hydrogen present higher risk (0.38-0.45) but potentially higher returns as costs decline.

Investors should consider technology diversification to balance mature, lower-risk assets with emerging, higher-potential investments. The correlation analysis suggests relatively low correlations between solar and wind returns due to differing resource availability patterns, supporting the case for combined portfolios. Energy storage investments can provide additional diversification benefits while enabling higher renewable penetration.

Geographic diversification across regulatory jurisdictions reduces policy risk exposure. Our case studies demonstrate substantial differences in policy stability and investment conditions across countries, suggesting that international diversification can enhance portfolio resilience. However, geographic diversification should be informed by careful assessment of local policy environments, grid integration conditions, and market structures.

\subsubsection{Risk Management Framework}

Infrastructure energy investments face multiple risk categories that require systematic management approaches. Policy risk emerges as particularly significant given the sensitivity of returns to carbon pricing and support mechanisms. Investors should conduct scenario analysis incorporating different policy pathways, engage in policy advocacy to support stable regulatory frameworks, and consider policy risk insurance products where available.

Technology risk can be managed through phased investment approaches, performance guarantees from equipment suppliers, and partnerships with technology developers. The rapid cost declines observed in solar PV and batteries illustrate both the upside potential of technological progress and the risk of technology obsolescence for earlier-vintage investments.

Market risk relating to electricity prices and offtake agreements requires careful contract structuring. Long-term power purchase agreements (PPAs) with creditworthy counterparties can provide revenue certainty, though they may limit upside participation in rising market prices. The optimal balance depends on investor risk preferences and portfolio composition.

Integration and curtailment risks are increasing as renewable penetration grows. Investments in grid infrastructure, energy storage, and demand response capabilities can mitigate these risks while generating additional revenue streams. Co-location of generation with storage or flexible demand offers particular promise.

\subsubsection{Long-term Value Investment}

The scenario simulations underscore the long-term value creation potential of clean energy investments aligned with climate objectives. Under the Ambitious scenario, cumulative investment of 150 trillion USD through 2050 generates estimated NPV of 45 trillion USD through avoided climate damages, energy cost savings, and co-benefits. Investors with long-term horizons should consider allocating capital to infrastructure energy investments as part of a climate-aligned portfolio strategy.

Impact measurement and reporting can enhance accountability and attract capital from sustainability-focused investors. Standardized metrics for carbon emissions avoided, renewable energy generated, and jobs created enable comparison across investments and demonstrate contribution to sustainable development objectives.

\subsection{International Cooperation and Sustainable Development}

The global nature of climate change and the interconnected structure of energy markets necessitate international cooperation to maximize the effectiveness of infrastructure energy investment.

\subsubsection{Technology Transfer and Capacity Building}

Accelerating clean energy deployment in developing economies requires substantial technology transfer and capacity building support from developed countries. Our analysis indicates that developing economies face higher capital costs, weaker policy frameworks, and limited technical capacity that constrain investment. International climate finance commitments, including the 100 billion USD annual target, should prioritize infrastructure energy investment with emphasis on capacity building and institutional strengthening.

South-South cooperation, particularly involving China's manufacturing capabilities and deployment experience, can complement North-South flows. However, such cooperation should be designed to build local capabilities rather than create long-term dependencies, with emphasis on training, technology licensing, and joint venture arrangements.

\subsubsection{Global Carbon Market Integration}

The emergence of carbon pricing mechanisms across jurisdictions creates opportunities for market linkage that can enhance efficiency and reduce leakage concerns. Article 6 of the Paris Agreement provides a framework for international carbon market cooperation, though implementation details remain under negotiation. Linked carbon markets can equalize marginal abatement costs across jurisdictions, support technology transfer through crediting mechanisms, and generate revenue for climate finance.

However, market linkage requires careful attention to environmental integrity, including robust accounting frameworks, baseline methodologies, and monitoring systems. The experience with the Clean Development Mechanism under the Kyoto Protocol illustrates both the potential and the pitfalls of international carbon crediting.

\subsubsection{Alignment with Sustainable Development Goals}

Infrastructure energy investment should be designed to contribute to multiple Sustainable Development Goals (SDGs) beyond climate action (SDG 13). Our case studies reveal substantial co-benefits for affordable and clean energy (SDG 7), decent work and economic growth (SDG 8), industry innovation and infrastructure (SDG 9), and sustainable cities and communities (SDG 11).

Systematic assessment of SDG impacts can inform project design and stakeholder engagement while demonstrating the broader value proposition of clean energy investment. Multi-criteria analysis frameworks that incorporate economic, environmental, and social dimensions can support more holistic investment decision-making consistent with sustainable development objectives.

The energy transition presents a historic opportunity to achieve climate objectives while advancing economic development and social inclusion. Realizing this potential requires coordinated action by governments, investors, and international institutions, guided by the empirical evidence and strategic frameworks presented in this analysis.

\section{Conclusion, Limitations, and Future Research}

This paper examined infrastructure energy investment as a strategic component of the low-carbon transition, emphasizing the interaction between policy design, technological conditions, and economic outcomes. By integrating quantitative panel data analysis, detailed case studies, and scenario simulations, we have examined the complex interplay of policy, technology, and economic factors that determine the effectiveness of energy infrastructure investments in achieving both economic growth and environmental sustainability objectives.

\subsection{Summary of Key Findings}

Our research addresses four core research questions, yielding the following principal findings.

Regarding the influence of infrastructure energy investment on global energy transition and low-carbon development, our panel data analysis demonstrates a statistically significant positive relationship between renewable energy investment and GDP growth (coefficient = 0.285, $p < 0.01$), alongside a significant negative relationship with carbon emissions (coefficient = -0.142, $p < 0.01$). These findings indicate that infrastructure energy investment can simultaneously advance economic growth and decarbonization objectives, challenging the conventional view that environmental protection necessarily involves economic trade-offs. The global renewable energy share increased from 18.2\% in 2010 to 34.8\% in 2023, with investment reaching 835 billion USD annually, demonstrating the substantial momentum of the energy transition.

Concerning the optimization of investment portfolios under policy guidance, our case studies reveal that effective policy frameworks combining long-term certainty, carbon pricing, and technology support can significantly enhance investment returns while achieving environmental objectives. The German Energiewende demonstrates the effectiveness of feed-in tariffs and sustained policy commitment, while the US Inflation Reduction Act illustrates the transformative potential of comprehensive industrial policy. Our ablation study quantifies the contribution of individual policy components: carbon pricing accounts for 31.3\% of investment NPV improvement, green finance mechanisms contribute 24.3\%, and technology support adds 28.2\%. The full integrated framework achieves 92.5\% policy effectiveness compared to 45.2\% for carbon pricing alone.

On the role of technological innovation in enhancing long-term returns, our analysis documents substantial cost reductions and performance improvements across clean energy technologies. Solar PV levelized costs declined from 125 USD/MWh in 2015 to 38 USD/MWh in 2023, while internal rates of return increased from 8.5\% to 16.8\% over the same period. Energy storage investments show particularly rapid improvement, with IRR increasing from 5.2\% to 15.8\% as battery costs declined by 68\%. The technology innovation index emerges as a significant predictor across all regression specifications, underscoring the importance of continued R\&D investment and deployment policies to maintain cost reduction trajectories.

Regarding risk-return evaluation and multi-objective optimization, our investment analysis reveals substantial heterogeneity across technologies and regions. Mature technologies (solar PV, onshore wind) offer risk scores of 0.15-0.18 with IRR of 16-19\%, while emerging technologies (hydrogen, offshore wind) present higher risk (0.25-0.45) but potentially superior returns as technologies mature. The sensitivity analysis identifies carbon price as the most influential parameter (sensitivity index = 0.78), followed by discount rate (0.75) and energy prices (0.71), providing actionable guidance for investment decision-making under uncertainty.

\subsection{Research Contributions}

This study makes several contributions to the academic literature and practical understanding of infrastructure energy investment.

First, we develop and empirically validate an integrated investment framework that systematically incorporates policy, technology, and economic factors. Unlike previous studies that typically examine these dimensions in isolation, our framework captures the synergistic effects and interdependencies that characterize real-world energy investment decisions. The ablation study methodology provides a rigorous approach for decomposing the contributions of individual framework components.

Second, we provide actionable, data-driven guidance for both policymakers and investors. The policy recommendations regarding carbon pricing trajectories, green finance innovation, and technology support are grounded in empirical evidence from our quantitative analysis and case studies. The investment strategy recommendations address portfolio optimization, risk management, and long-term value creation in ways that can directly inform capital allocation decisions.

Third, we contribute to methodological innovation by combining panel econometrics, comparative case studies, and scenario simulation within a unified analytical framework. This multi-method approach enhances the robustness and generalizability of our findings while providing complementary perspectives on the research questions.

\subsection{Limitations}

Despite these contributions, our study has several limitations that should be acknowledged.

First, while our panel dataset covers 15 countries representing the majority of global renewable investment, it may not fully capture the diversity of conditions in smaller or less-developed economies. The generalizability of our findings to countries with substantially different institutional contexts, resource endowments, or development priorities warrants further investigation.

Second, the scenario simulations rely on simplifying assumptions regarding technological progress, policy evolution, and economic growth that may not materialize in practice. The scenarios should be understood as illustrative explorations of possible futures rather than forecasts. Actual outcomes will depend on complex political, social, and technological dynamics that are inherently difficult to predict.

Third, our analysis focuses primarily on economic and environmental outcomes, with limited attention to social dimensions such as distributional effects, employment quality, and community impacts. A more comprehensive assessment would integrate social impact analysis to provide a fuller picture of the sustainability implications of infrastructure energy investment.

Fourth, data availability constraints limit our ability to examine certain important variables, particularly at the project level. More granular data on individual investment characteristics, financing terms, and operational performance would enable more precise estimation of relationships and mechanisms.

\subsection{Future Research Directions}

Building on this study, several promising avenues for future research emerge.

First, extending the analysis to a broader range of countries, particularly developing economies and emerging markets, would enhance the generalizability of findings and provide insights into the specific challenges and opportunities these contexts present. Comparative analysis of energy transition pathways across different development stages and institutional settings could yield valuable policy lessons.

Second, deeper investigation of the mechanisms linking infrastructure energy investment to economic outcomes would strengthen the causal interpretation of our findings. Natural experiments arising from policy changes, combined with high-frequency data and quasi-experimental methods, could provide more precise identification of causal effects.

Third, integration of social impact assessment into investment evaluation frameworks represents an important frontier. Developing standardized methodologies for measuring and comparing social outcomes across projects would enable more holistic decision-making consistent with sustainable development objectives.

Fourth, the emerging role of digital technologies, artificial intelligence, and advanced analytics in optimizing energy system operations and investment decisions warrants systematic investigation. Smart grid technologies, predictive maintenance, and demand response capabilities may substantially alter the risk-return profiles of energy infrastructure investments.

Fifth, as the energy transition accelerates, understanding the dynamics of stranded assets and transition risks becomes increasingly important. Research on optimal phase-out strategies for fossil fuel infrastructure, support mechanisms for affected communities and workers, and financial system resilience to climate-related risks would provide valuable guidance for policymakers and investors.

In conclusion, infrastructure energy investment stands at the intersection of some of the most significant challenges and opportunities of our time. The evidence assembled in this study demonstrates that with appropriate policy frameworks, technological innovation, and strategic investment approaches, the transition to a low-carbon energy system can generate substantial economic, environmental, and social benefits. Realizing this potential requires sustained commitment and coordinated action by governments, investors, and civil society, guided by rigorous analysis and evidence-based policy design. We hope this research contributes to that essential endeavor.

\newpage
\bibliography{references}

\end{document}